\documentclass[aps,prb,twocolumn,preprintnumbers,showpacs,amsmath,amssymb,floatfix]{revtex4}
\usepackage{graphicx, bm}
\usepackage{float}
\usepackage{latexsym}
\usepackage{amsmath}
\usepackage{graphics}
\usepackage{amssymb}
\usepackage{layout}
\usepackage{color}
\usepackage{verbatim}
\usepackage{amsfonts,epsfig}

\newcommand{\beq}{\begin{equation}}
\newcommand{\eeq}{\end{equation}}
\begin{document}
\title{Collective modes of monolayer, bilayer, and multilayer fermionic dipolar liquid}
\author{Qiuzi Li, E. H.\ Hwang, and S. Das Sarma}
\affiliation{Condensed Matter Theory Center, Department of Physics,
  University of Maryland, College Park, Maryland 20742}
\date{\today}

\begin{abstract}
Motivated by recent experimental advances in creating polar molecular gases in the laboratory, we theoretically investigate the many body effects of two-
dimensional dipolar systems with the anisotropic and $1/r^3$ dipole-
dipole interactions. We calculate collective modes of 2D dipolar
systems, and also consider spatially separated bilayer and multilayer superlattice
dipolar systems. We obtain the characteristic features of collective
modes in  quantum dipolar gases. We quantitatively compare the
modes of these dipolar systems with the modes of the extensively studied usual
two-dimensional electron systems, where the inter-particle interaction is Coulombic.
\end{abstract}
\maketitle

\section{Introduction}
Recent experimental progress in producing and manipulating ultracold
polar molecules with a net electric
dipole moment\cite{Ospelkaus_NP08,Ni_Sci08,Ni_PCCP09,Ospelkaus_PRL10},
provides new possibilities to explore novel quantum many-body physics
in such systems\cite{Griesmaier_PRL77,Ospelkaus_FD09,Mingwu_PRL10,McClelland_PRL06}.
A series of theoretical work have been done within such
dipolar system, such as the stable topological $p_x+i p_y$ superfluid
phase created with fermionic polar molecules with large dipole moment
confined to a two dimensional geometry\cite{Cooper_PRL09}, the
existence of spontaneous interlayer superfluidity in bilayer systems
of cold polar molecules\cite{Roman_arX09}, the anisotropic Fermi
liquid theory for the ultra cold fermionic polar
molecules\cite{Chan_PRA10}, the zero sound mode in three dimensional
(3D) dipolar Fermi gases\cite{Ronen_arX09}, the superfluid properties
of a dipolar Bose-Einstein condensate (BEC) in a fully three
dimensional trap\cite{Wilson_PRL10} and the finite temperature
compressibility of a fermionic dipolar gas\cite{Jason_PRA10}.

Most physical systems have long-lived excited states by conserving the
total number of particles. These excitations have a bosonic
character and are known as collective modes.
A collection of charged particles
is characterized by a collective mode associated with
the self-sustaining in-phase density oscillations of all the
particles due to the restoring force on the displaced
particles, which arises from the self-consistent electric field
generated by the local excess charges \cite{Mahan}.
A two dimensional (2D) charged system can support
density oscillations with the
long-wavelength dispersion $\omega \propto \sqrt{q}$.
A similar collective mode occurs in a neutral
Fermi system\cite{Fetter}. A repulsive short-range interparticle potential is
sufficient to guarantee such a mode. This resulting density
oscillation turns out to have a linear dispersion relation $\omega
\propto q$ in the long wavelength limit and is known as zero sound.
The zero sound mode in connection with the RPA linear response theory
was originally discovered by Landau as a collective oscillation of the
Fermi liquid with short-range inter-particle interaction as appropriate
for a neutral system. Thus
the zero sound and plasma oscillation are physically very
similar, both being collective modes of an interacting Fermi liquid.  However,
the zero sound is physically very different from
ordinary first sound,
despite the similar dispersion relations $\omega \propto q$.

In this paper we investigate the collective mode
of the dipolar system, which has
anisotropic and  $1/r^3$ dipolar interaction instead of the isotropic
$1/r$ Coulomb
interaction. The collective  mode of 3D dipolar
systems is recently discussed as a solution of the
linearized Boltzmann equation \cite{Chan_PRA10,Ronen_arX09}. However,
our theory is based on the leading order expansion of
the dynamically screened dipolar interaction, the
so-called infinite bubble diagram expansion, with each bubble being
the noninteracting irreducible polarizability.
In our approach, the detailed form of the long-wavelength dispersion is
fixed by the behavior of $q^2 V(q)$ as $q \rightarrow 0$, where $V(q)$
is the dipolar interaction in momentum space. Since $V(q)$ is
anisotropic and behaves as
a short-range potential we have the very interesting and zero sound
like-collective modes in the long wavelength limit.
We also consider a double layer dipolar system formed by two parallel
single-layer
dipolar systems separated by a distance $a$ and a multilayer dipolar superlattice
made of periodic arrays of 2D dipolar
systems in the direction transverse to the 2D plane.
Collective modes of 2D multi-layer structures have been extensively
studied since the existence of an undamped acoustic plasmon mode
was predicted in semiconductor double quantum well systems
\cite{Sarma_PRB81}.
The collective modes can be detected with
experimental probes that couple directly to the particle density
operators. Typical experiments for solid state systems are
inelastic light scattering spectroscopy\cite{Abstreiter,Pinczuk_PRL86,Olego_PRB82,Eriksson_PRL99}, frequency-domain far-infrared\cite{Allen_PRL77} or
microwave spectroscopy, or inelastic electron-scattering
spectroscopy\cite{Liu_PRB08,Liu_PRB10,Langer_NJP10,Karmberger_PRL08,Lu_PRB09,Eberlein_PRB08}.

The layout of the paper is as follows: In Sec. \ref{sec:single}, we
derive both analytical formula and numerical results for the plasmon
dispersion relation in the 2D monolayer dipolar gas. In
Sec. \ref{sec:bilayer}, we study the plasmon modes in bilayer dipolar
gas and their loss function (spectral strength). In
Sec. \ref{sec:sup}, we present analytical results of the plasmon modes
in the dipolar superlattice system within a simple model. In Appendix
\ref{app:ftinterlayer} and \ref{app:superla}, we provide the detailed
calculation for the interlayer dipolar interaction and two summations
in Sec.\ref{sec:sup}, respectively.

\section{collective mode in monolayer  dipolar system}
\label{sec:single}

We start from the fundamental many-body formula defining the
collective mode of a fermionic dipolar system.
The collective mode of a fermionic dipolar system is
given by the dynamical structure factor	$S(q,\omega)$,
which is proportional to Im[$\epsilon(q,\omega)^{-1}$],
where $\epsilon(q,\omega)$ is the dynamical dielectric function
of the system.
The longitudinal collective-mode dispersion
can be calculated by looking for poles of the density
correlation function, or equivalently, by looking for zeros of the
dynamical dielectric function.
\begin{equation}
\epsilon(q,\omega)=1-V(q) \Pi_0(q,\omega),
\end{equation}
where $q$ and $\omega$ are, respectively,  the 2D wave
vector parallel to the plane and the frequency,
$V(q)$ is the interaction between dipolar molecules in wave-vector
space, and $\Pi_0(q,\omega)$ is the leading-order irreducible
polarizability (i.e., the so-called bare bubble
or the Lindhard function in the relevant dimension).

\begin{figure}
\includegraphics[width=0.86\columnwidth]{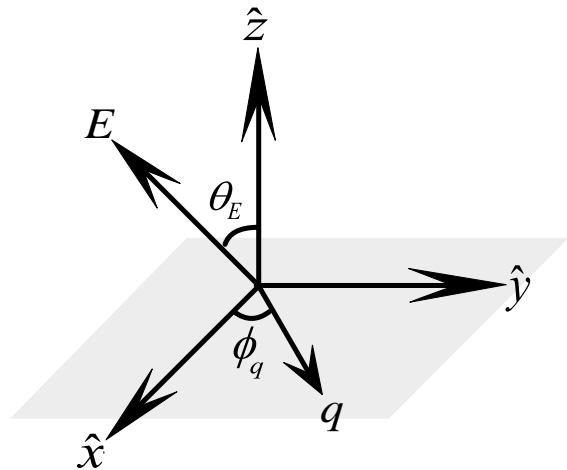}
\caption{(Color online). Schematic sketch of the 2D monolayer dipolar
  system in momentum space. The polarization direction of dipolar
  molecules is controlled by the external electric field  $\vec{E}$,
  which is in the $xz$ plane.}
\label{fig:2D}
\end{figure}

The interaction between the dipolar molecules is spatially anisotropic,
which depends not only on the distance between two dipole molecules
but also the angle between their relative vector and dipolar
orientations. For the dipolar system in a 2D
plane ($xy$-plane), the interaction between two dipoles located at
$r_1$ and $r_2$, respectively, within the layer can be written as:
\begin{equation}
V_{2D} (\vec{r_1}-\vec{r_2})=\frac{d^2}{|\vec{r_1}-\vec{r_2}|^3}(1-3
\sin^2\theta_E \cos^2\phi)
\label{Eq:s2d}
\end{equation}
where $d$ is the electric dipole moment. If the external electric
field $\vec{E}$
is set in the $xz$-plane with an polar angle of $\theta_E$, then
$\phi$ is the azimuthal angle relative to the $x$-axis. The
configuration is depicted in Fig. \ref{fig:2D}.
After Fourier transformation, we could get the dipolar interaction in
the wave vector space. In order to handle with the short distance
divergence of the 2D dipolar interaction and since we are
more interested in the long wave length limit, we use the short
distance cutoff $c$ beyond which the dipolar interaction, given by
Eq. \ref{Eq:s2d}, is valid\cite{Chan_PRA10}.
For the dilute Fermi gas, the short distance cut-off is set to satisfy
the relation $k_F c \ll 1$. Then we have the interaction in wave
vector space
\begin{equation}
V_{2D} (q)= 2\pi d^2 P_2(\cos \theta_E)(\frac{1}{c}-q)+\pi d^2 q \sin^2
\theta_E \cos 2 \phi_q,
\label{Eq:2D}
\end{equation}
where $P_2(\cos \theta_E)$ is the second Legendre
polynomial. Alternatively, if we start from the 3D
dipolar interaction and assume a fixed
Gaussian density profile in the $z$ direction $n (k_z) = e^{-k_z^2 w^2/4}$,
where $w$ characterizes the typical confinement size of the two
dimensional bipolar system in the $z$ direction,
then integrating over the $z$ direction, as shown in the Eq. 3 of
Ref.~[\onlinecite{Jason_PRA10}], yields the same result with a
numerical factor of order unity in front of $1/w$. From
Eq.~(\ref{Eq:2D}), we can see that the first term of the $V_{2D} (q)$ is
isotropic, which can be either positive or negative depending on the
direction of the external electric field. While the second term of
Eq.~(\ref{Eq:2D}) is the anisotropic component, which can also be either
positive or negative but depending on the direction of $q$.

The leading order irreducible polarizability is given by
the bare bubble diagrams \cite{Mahan}  and for the spinless fermions
we have
\beq
\Pi({\bf q}, \omega) = \int \dfrac{d^2 k}{(2 \pi)^2}\dfrac{n_F
  (\xi_{\bf k})-n_F (\xi_{\bf k+q})}{\omega+\xi_{\bf k}-\xi_{\bf
    k+q}+i \eta}
\label{Eq:p1}
\eeq
where $\xi_{\bf k} = \hbar^2 k^2/2m$ is
the energy of single particle and $n_F$ is the Fermi distribution
function. At zero temperature, the polarizability was calculated by
Stern\cite{Stern_PRL67} and the exact expressions, written in terms of
dimensionless parameters $x =q/2k_F$ ($k_F$ is the Fermi wave vector)
and $\delta = \omega/4E_F$ ($E_F$ is the Fermi energy), is given by
\cite{Victor_PRB04}:
\begin{equation}
\begin{array}{l l l }
\text{Re} \Pi(x, \delta) = \dfrac{m}{2\pi x}
\Big [ \text{sgn}(x-\frac{\delta}{x}) \sqrt{(x-\frac{\delta }{x})^2-1}
\\
\ \ \ \ \ \ \ \ \ \ \ \ \ \ \ \ \ \ \ \ \ + \sqrt{(x+\frac{\delta}{x})^2-1}-1 \Big]
\\
\\
\text{Im} \Pi(x, \delta) = \dfrac{m}{4\pi x}
\Big[\sqrt{1-(x+\frac{\delta
    }{x})^2}-\sqrt{1-(x-\frac{\delta }{x})^2} \Big]
\label{Eq:Pola}
\end{array}
\end{equation}
where $\text{Re} \Pi(x, \delta)$ and $\text{Im} \Pi(x, \delta)$ are
the real part and the imaginary part of the polarizability, respectively.

We could further introduce some dimensionless parameters which we will
use in the following discussions: $s= \frac{m d^2}{2
  \hbar^2 c }$ and $r = 2 d^2 k_F m/ \hbar^2$ which describes the
strength of dipolar
interaction. For example, experimental parameters of fermionic
$^{40}$K$^{87}$Rb ($m = 127$ amu,$d= 0.57 $ Debye, $n =
10^{8}$cm$^{-2}$), we have $c = 10$ nm ($c k_F \sim 0.03 \ll 1$), $s
\simeq 31$ ($s \gg 1$) and $r=4.4$. We use these parameters in the
numerical calculation.

For a single layer fermionic dipolar system, the collective modes
can be found by looking for the zeros of the dynamical dielectric
function, i.e.,
\beq
\epsilon (q, \omega) = 1- V(q) \Pi(q, \omega) = 0.
\label{Eq:sd}
\eeq
First, we consider the
leading order wave vector dependence of the collective mode. In the
long wavelength limit ($\omega \gg q v_F$, where $v_F=k_F/m$ is the
Fermi velocity), the 2D polarizability becomes
\begin{equation}
\Pi(q, \omega)= \alpha \frac{q^2}{\omega^2}+O( \frac{q^4}{\omega^4}),
\label{Eq:P}
\end{equation}
where $\alpha={n}/{m}$, ($n$ is the
2D density of the dipolar molecules and is related to
the Fermi wave vector $k_F$ as $k_F = \sqrt{ 4
  \pi n}$). Then, we have the collective mode in the long wave length
($q \rightarrow 0$)
\begin{equation}
\omega(q) \simeq q \sqrt{(2\pi d^2 P_2(\cos \theta_E)(\frac{1}{c}-q)+\pi
  d^2 q \sin^2 \theta_E \cos 2 \phi_q) \dfrac{n}{m}}
\label{Eq:single}
\end{equation}

As $\theta_E$ varies from 0 to $\frac{\pi}{2}$
by changing the direction of external field,
the dipolar interaction changes from an isotropic
repulsive to an attractive one at large value of $\theta_E$. The
plasmon mode given in Eq.~(\ref{Eq:single}) depends on both the
direction of the external field and the direction of the 2D wave
vector. Since the 2D dipolar interaction in the wave
vector space has both s-wave and d-wave symmetry as shown in
Eq.~(\ref{Eq:2D}), we only need to consider the case with $\phi_q$ in
the range $[0, \frac{\pi}{2}]$. As special cases we investigate the
plasmon modes along the $x$-axis and $y$-axis, which corresponds to $\phi_q =
0$ and $\phi_q = \frac{\pi}{2}$. For
$\phi_q = 0$, we have the plasmon mode at long wave
length limit from Eq.~(\ref{Eq:single}):
\beq
\omega \simeq q \sqrt{(2\pi d^2 P_2(\cos \theta_E)(\frac{1}{c}-q)+\pi
  d^2 q \sin^2 \theta_E ) \dfrac{n}{m}}.
\eeq
We find from the above formula that when the electric field is
perpendicular to the $xy$ plane (i.e. $\theta_E = 0$), the plasmon mode
becomes
\beq
\omega (q \rightarrow 0) \simeq q\sqrt{\dfrac{2 \pi d^2 n}{c \ m}}.
\label{Eq:p2D}
\eeq
Using $k_F = \sqrt{4 \pi n}$ and $s = \dfrac{m d^2}{2 \hbar^2 c }$ we
have
\beq
\omega (q \rightarrow 0) \simeq  v_F \sqrt{s} q.
\label{Eq:siso}
\eeq
Since $s \gg 1$, the plasmon mode shown in Eq. \ref{Eq:p2D} satisfies
the consistency
criterion $\omega \gg q v_F$, which has been used to get the 2D
polarizability up to the leading order in wave vector.
Note that for $\omega > v_F q$ the mode lies above the single
particle excitation (SPE) regime (or particle-hole continuum)
and prevents its direct decay through coupling to
particle-hole continuum.
The SPE region is defined by the nonzero value of the
imaginary part of the total dielectric function, Im$[\epsilon (q,
  \omega)] \neq 0$, which gives rise to the damping of a plasmon mode
by emitting a particle-hole pair excitation\cite{Euyheon_PRB09}.

When the direction of electric field satisfies $3 \cos^2 \bar{\theta}
=1$ (or $\bar{\theta}
\simeq 55^\circ$), the short distance cut-off
disappears and the plasmon dispersion relation becomes
\begin{equation}
\omega \simeq q^{\frac{3}{2}} \sqrt{\dfrac{2 \pi d^2 n}{3m}} =
\sqrt{\dfrac{r q}{12 k_F}} v_F q.
\label{Eq:sc2d}
\end{equation}
We see  that the undamped plasmon mode in Eq.~(\ref{Eq:sc2d})
exists only for $q  > q_c=12 k_F/r$. For $q < q_c$ the mode enters
into the single particle excitation region and it is damped by
producing particle-hole pair.
As the external electric
field is further tilted leading to $3 \cos^2 \theta_E < 1$
the solution for $\omega$ satisfying
Eq.~(\ref{Eq:sd}) is purely imaginary, and there  is no well-defined
collective mode. Thus, along $x$ axis, the direction of the external
electric field must be smaller than the critical direction $\theta_c$
in order to exist an undamped plasmon mode.

\begin{figure}
\includegraphics[width=0.99\linewidth]{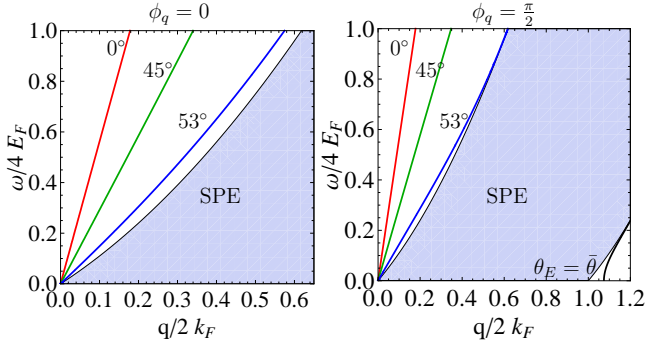}
\caption{(Color online). Calculated plasmon mode dispersions of single
  layer dipolar system for different external electric field direction
  $\theta_E$.  The shadowed region indicate the single-particle
  excitation (SPE) region. The interaction parameter $r=4.4$ and $s =
  31$ are used in this calculation.
 The mode dispersions (a) for $\phi_q = 0$ and
(b) for $\phi_q = \dfrac{\pi}{2}$ are shown. }
\label{fig:S}
\end{figure}

Now we consider the collective mode along the $y$-axis (i.e. $\phi_q =
\frac{\pi}{2}$). The plasmon mode for this case derived from
Eq.~(\ref{Eq:single}) is given by
\begin{equation}
\omega(q) \simeq q \sqrt{(2\pi d^2 P_2(\cos \theta_E)(\frac{1}{c}-q)-\pi
  d^2 q \sin^2 \theta_E) \dfrac{n}{m}}
\label{Eq:sy}
\end{equation}
For an electric field being perpendicular to the single layer plane
$\theta_E = 0$, the plasmon mode of the system is
isotropic and it is the same as given in Eq.~(\ref{Eq:siso}). For $3 \cos^2
\theta_E = 1$, there is an undamped mode at $\omega
\ll q v_F$ and $q > 2k_F$ because Lindhard function becomes negative
and the dipolar interaction $V_{2D} (q)$ is
also negative along $y$-axis for $3 \cos^2 \theta_E = 1$.
To get this unusual mode we expand the 2D polarizability for $\omega
\ll  q v_F$ as
\beq
\Pi(x, \delta)=\dfrac{m}{2\pi}
\Big[\frac{\sqrt{-1+x^2}}{x}-1-\frac{\delta ^2}{2 \left(x^3
    \left(-1+x^2\right)^{3/2}\right)}\Big].
\eeq
Using this large $q$ behavior of the polarizability and
Eq.~(\ref{Eq:sd}) we have the undamped low energy collective mode
for $\phi_q =\frac{\pi}{2}$ and $\theta_E = \bar{\theta}$, which is consistent
with the numerical result shown in the left panel of Fig.~\ref{fig:S}.
\begin{equation}
\omega  = \frac{v_F q}{2\sqrt{2} \sqrt{r}}
\left(-4+\frac{q^2}{k_F^2}\right)^{3/4} \sqrt{6-\frac{q
    r}{k_F}+\sqrt{-4+\frac{q^2}{k_F^2}} r}.
\end{equation}
The low energy mode at large wave vectors does
not exist in the usual two dimensional electron system (2DES)
since the interaction of 2DES is isotropic and positive. This mode is
unique for a fermionic dipolar system.
Compared with the 2D fermionic dipolar system, the long wave-length plasma frequency for the extensively studied 2DES is written as\cite{Sarma_PRL09}
\beq
\omega_2(q \rightarrow 0)=\sqrt{\dfrac{2\pi n e^2}{\kappa m}}q^{1/2}+O(q^{3/2})
\eeq
where $n$ and $m$ are the charge carrier density and the effective mass of the charge carriers in the 2DES, respectively. The plasma frequency for the 2DES is isotropic and proportional to $\sqrt{q}$, while the plasma frequency of the 2D dipolar system is characterized by Eq. \ref{Eq:single}, which is anisotropic and has two different dispersions at long wave-length limit.

In Fig.~\ref{fig:S}, we show numerically calculated collective
mode dispersions of single layer dipolar system for three different
external electric field directions $\theta_E$. The calculated numerical
results agree well with the analytical results discussed above. We choose
$\theta_E = 53^\circ$ (it can be any angle close to the critical value
$\bar{\theta}$)
to investigate the behavior of the collective mode near to the
critical angle $\theta_E = \bar{\theta}$.

When the 2D wave vector is along the $x$-axis ($\phi_q = 0$),
the slope of plasmon dispersion decreases as the electric field
direction increases. In particular,
for $ \theta_E>\bar{\theta} \approx 55^\circ$, the plasmon mode enters
into the SPE region and it is overdamped by producing particle-hole
pair. On the other hand,
when the 2D wave vector along the $y$-axis ($\phi_q =
\frac{\pi}{2}$), the plasmon mode has similar feature as the case for
$\phi_q = 0$ and for smaller angle of $\theta_E$. When $\theta_E$
approaches to the
critical value $\bar{\theta}$ (i.e., $\bar{\theta} \simeq 55^\circ$),
the long-wave length plasmon mode approaches to the upper boundary of
the SPE region (i.e., $\omega = {q^2}/{2 m}+v_F q$).
The main difference between $\phi_q=0$ and $\phi_q=\pi/2$ cases is
that there is a short-wave length plasmon mode for the latter case
even  at $ 3 \cos^2 \theta_E \leq 1$. This
plasmon mode for $\phi_q=\pi/2$ appears below the lower boundary
of the SPE region (i.e., $\omega = {q^2}/{2 m}-v_F q$), where both the
dipolar interaction and the Lindhard function are negative.

\section{Plasmon mode in Bilayer dipolar system}
\label{sec:bilayer}

For a bilayer system,
which is parallel to the $xy$
plane and is separated by a distance $a$,
we need to consider
the generalized dielectric tensor $\epsilon$ \cite{Sarma_PRB81} in order
to find the collective modes.
Within the RPA the $lm$ component of the dielectric tensor is given by
\beq
\epsilon_{lm}(q,\omega) = \delta_{lm}-V_{lm}(q)\Pi_m(q,\omega),
\eeq
where $l,m =1$ or 2
with 1,2 denoting the index of two different layers. Then
the plasmon modes are given by the zeros of the
two-component determinantal equation, i.e.,
\begin{eqnarray}
\begin{array}{l c c c c c c}
\epsilon (q , \omega) = [1- V_{11}(q) \Pi_1 (q , \omega)][1- V_{22}(q)
  \Pi_2 (q , \omega)]
\\
\\
\ \ \ \ \ \ \ \  - V_{12}(q) V_{21}(q) \Pi_1 (q , \omega) \Pi_2 (q ,
\omega) =0,
\label{Eq:D}
\end{array}
\end{eqnarray}
where $\Pi_l$
is the polarizability of layer $l$ and is given by Eq.~(\ref{Eq:p1}),
and $V_{ll}$ and $V_{lm}$ are, respectively, the intralayer and
interlayer dipolar interaction.
The intralayer dipolar interaction is the same as given in
Eq.~(\ref{Eq:2D})
\begin{eqnarray}
\begin{array}{l c c c c c c}
V_{11}(q)= V_{22}(q)= 2\pi d^2 P_2(\cos \theta_E)(\dfrac{1}{c}-q)
\\
\ \ \ \ \ \ \ \ \ \ \ \ +\pi d^2 \sin^2 \theta_E  \cos 2 \phi_q q
\\
\ \ \ \ \ \ \ \ = \pi d ^2 (3 \cos^2 \theta_E - 1) \dfrac{1}{c}
\\
\ \ \ \ \ \ \ \ \ \ \ \ + 2 d^2 q \pi (\sin ^2 \theta _E\cos ^2 \phi
_q-\cos ^2 \theta _E).
\label{Eq:intra}
\end{array}
\end{eqnarray}

For the 2D bilayer dipolar system
the interlayer interaction $V_{12}$ is given in real space
\begin{equation}
V_{12}(\vec{r})=\frac{d^2}{(r^2+a^2)^{\frac{3}{2}}}\left[1-\frac{3(r
    \cos\phi \sin\theta_E+h \cos\theta_E)^2}{r^2+h^2}\right],
\end{equation}
where $\theta_E$ is the polar angle of
an external electric field which is applied in the $xz$ plane and
$\phi_q$ is the azimuthal angle measuring from $x$-axis.
Note  that $V_{12}(-\vec{r}) \neq
V_{12}(\vec{r})$ if the electric field is not perpendicular to the
$xy$ plane, which gives rise to the imaginary component for the interlayer
interaction in the wave vector space. The detailed calculation for
$V_{12}(q)$ is given in Appendix. \ref{app:ftinterlayer} and we express
the following explicit form of the interlayer interaction in momentum space as:
\begin{eqnarray}
\begin{array}{l c c c c c c}
V_{12}(q) =  2 d^2 q \pi  e^{-a q} (\sin ^2 \theta _E\cos ^2 \phi _q-\cos ^2 \theta _E
\\
\\
\ \ \ \ \ \ \ \ \ \ \ \ -\sin  2\theta_E\cos  \phi _q \ i)
\label{Eq:inter1}
\end{array}
\end{eqnarray}
and
\begin{eqnarray}
\begin{array}{l c c c c c c}
V_{21}(q)=  2 d^2 q \pi  e^{-a q} (\sin ^2 \theta _E\cos ^2 \phi _q-\cos ^2 \theta _E
\\
\\
\ \ \ \ \ \ \ \ \ \ \ \ + \sin  2\theta_E\cos  \phi _q \ i)
\label{Eq:inter2}
\end{array}
\end{eqnarray}

We note that both the interlayer and intralayer interaction depends on
both the direction of momentum $q$ and the direction of an electric field
$\vec{E}$. In
addition, there is a short distance cutoff in interlayer interaction,
Eq.~(\ref{Eq:intra}), but not in the intralayer interaction.
The interlayer distance $a$ should be
larger than the short distance cutoff $c$ beyond which the interaction
between two dipole molecules can be described by the dipolar
interaction. Typically in the cold atomic system, the interlayer
distance $a = 500$nm and we use $c = 10$nm as the short distance
cut-off for the intralayer interaction as done in the single layer
dipolar system. Alternatively,  we can assume that the
two layers separated by a distance $a$ are strongly confined in the
$z$ direction with Gaussian distribution in the
$z$ direction, $n (z) \propto e^{{-(z\pm a/2)^2}/{2 w^2}}$.
In the wave vector space we have $n(k_z) =
e^{-{w^2 k_z^2}\pm i{d}/{2} k_z}$.
Then, we get the same interlayer interaction when we
integrate out the $k_z$ part of
the 3D dipolar interaction.
The numerical factor before $1/w$ is
suppressed by $e^{-\frac{a^2}{2w^2}}$ of Eq. (1) in
Ref.~[\onlinecite{Jason_PRA10}], which means that the term with $1/w$
can be neglected as long as $a/w \gg 1$. We numerically checked that
the interlayer interaction with Gaussian distribution agree well with
Eq.~(\ref{Eq:inter1}) as long as $a/c>5$, which is the case we
consider in our work.

\subsection{Long wavelength plasmon mode in bilayer dipolar system}

In this subsection, we derive the analytical results of plasmon mode
in the long wave length limit.
We first consider the leading order wave
vector dependence of the plasmon mode in the bilayer dipolar gas. At
zero temperature, the two dimensional non-interacting polarizability
has the following limiting forms in the high frequency regimes (i.e.,
$\omega \gg q v_F$)\cite{Sarma_PRB81}:
\begin{equation}
\Pi_i(q, \omega)= \alpha_i \frac{q^2}{\omega^2}+O( \frac{q^4}{\omega^4})
\label{Eq:P2}
\end{equation}
where $i =1$ or 2 with 1,2 denoting the two different layers, and
$\alpha_i={n_i}/{m}$, $n_i$ is the dipolar
molecule density of the $i$th layer.
Then, combining Eq.~(\ref{Eq:D},\ref{Eq:intra}) and
Eqs.~(\ref{Eq:inter1})-(\ref{Eq:P2}), we obtain the
long-wavelength plasmon modes of the bilayer dipolar system:
\begin{eqnarray}
\begin{array}{l l }
\omega^2_\pm \simeq \dfrac{q^2}{2} \big(\alpha_1 V_{11}+\alpha_2 V_{22}
\\
\\
\ \ \ \ \ \pm \sqrt{(\alpha _1 V_{11}-\alpha _2 V_{22})^2+4 \alpha _1
  \alpha _2 V_{12}V_{21}}\big),
\label{Eq:general}
\end{array}
\end{eqnarray}
where $\omega_{+}$ ($\omega_-$) indicates the optical (acoustic)
plasmon mode where
the density fluctuations in each layer
oscillate in-phase (optical) and out-of-phase (acoustic)
relative to each other, respectively. Eq. \ref{Eq:general} is also valid for the ordinary bilayer 2DES, while the interlayer and intralayer interaction is written as:
\begin{eqnarray}
\begin{array}{l c c c c c c}
V_{11}(q)=V_{22}(q)= \dfrac{2\pi e^2}{q}
\\
\\
V_{12}(q)=V_{21}(q) = \dfrac{2 \pi  e^2}{ q}  e^{-a q}.
\label{Eq:2des}
\end{array}
\end{eqnarray}
which are both isotropic. In the strong coupling ($q a \ll 1$) and long wave-length limit,  we have the following two plasma frequencies for the 2DES:
\begin{eqnarray}
\begin{array}{l c c c c c c}
\omega_{+} \simeq \sqrt{\dfrac{2\pi e^2 (n_1+n_2)}{\kappa m}}q^{1/2}
\\
\\
\omega_{-} \simeq \sqrt{\dfrac{2\pi  e^2 n_1 n_2}{\kappa m (n_1+n_2)}}q
\end{array}
\end{eqnarray}
While in the weak coupling ($q a \gg 1$) and long wave-length limit , the modes are simply the respective two-dimensional plasma frequencies of the two components\cite{Sarma_PRB81}:
\begin{eqnarray}
\begin{array}{l c c c c c c}
\omega_{+} \simeq \sqrt{\dfrac{2\pi n_1 e^2}{\kappa m}}q^{1/2}
\\
\\
\omega_{-} \simeq \sqrt{\dfrac{2\pi n_2 e^2}{\kappa m}}q^{1/2}
\end{array}
\end{eqnarray}

Now we investigate the plasmon modes of this bilayer dipolar
system in two regimes: the strong coupling limit ($q a \ll 1$) and the
weak coupling limit ($q a \gg 1$).

\subsubsection{Strong coupling limit $q a \ll 1$}

When the external
electric field is perpendicular to the bilayer plane, i.e., $\theta_E
=0$, the interlayer and intralayer
interaction become for $qa \ll 1$
\begin{eqnarray}
\begin{array}{l c c c c c c}
V_{11}(q)=V_{22}(q)= 2\pi d ^2 (\dfrac{1}{c}-q)
\\
\\
V_{12}(q)=V_{21}(q) = -2 \pi  d^2 q  e^{-a q}.
\label{Eq:dstong1}
\end{array}
\end{eqnarray}
Then the long wavelength plasmon modes become
\begin{equation}
\omega^2_\pm \simeq \dfrac{q^2 \pi d^2 }{c}\left[\alpha_1+\alpha_2 \pm
  \sqrt{(\alpha_1-\alpha_2)^2+ 4 q^2 c^2\alpha_1\alpha_2}\right].
\label{Eq:diso}
\end{equation}
Because the short distance cut-off in the intralayer dipolar
interaction is set to satisfy $k_{F_{1,2}} \ll \frac{1}{c}$ and
$r_{1,2} = 2 m d^2 k_{F_{1,2}}  $, it is clear to see that the above
two plasmon modes also satisfy the self-consistent criterion $\omega
\gg q v_F$. These two modes lies outside of the SPE region, which are
undamped and stable. Since the external electric field is
perpendicular to the $xy$ plane, the bilayer dipolar system is purely
isotropic, which gives rise to the plasmon mode to be independent of
the wave vector direction, $\phi_q$.

For $\theta_E =
\bar{\theta} \approx 55^\circ$, the system is
purely anisotropic and the plasmon modes are given by
\begin{eqnarray}
\begin{array}{l l l}
\omega^2_\pm \simeq \dfrac{d^2 \pi  q^3}{3}\Big[(\alpha_1+\alpha_2 )\cos 2\phi_q
\\
\\
\ \ \ \ \ \pm  \sqrt{(\alpha_1-\alpha_2)^2 \cos^2 2\phi_q+4
  \alpha_1\alpha_2(2+\cos 2\phi_q)^2}\Big].
\label{Eq:ds2}
\end{array}
\end{eqnarray}
When the 2D wave vector is along the $x$-axis (i.e., $\phi_q =0$),
the $\omega_-$ plasmon mode is overdamped
since $\omega_-$ becomes pure imaginary. But, the in-phase mode is
well defined as long as $r \gg 1$ (in order to satisfy the
self-consistent criterion $\omega \gg q v_F$)
and the plasmon mode is given by
\begin{eqnarray}
\begin{array}{l l l}
\omega^2_+ \simeq \dfrac{d^2 \pi  q^3}{3}\Big[(\alpha_1+\alpha_2
  )+\sqrt{(\alpha_1+\alpha_2)^2+32 \alpha_1\alpha_2}\Big].
\end{array}
\end{eqnarray}
However, along the $y$-axis ($\phi_q=\pi/2$), both
$\omega_-$ and $\omega_+$ plasmon modes are pure imaginary, and there
are no well-defined collective mode in the long wavelength limit.

\subsubsection{weak coupling limit $q a \gg 1$}

In weak coupling limit $e^{-a q}\approx
0$ and the interlayer
interaction between two layers is much weaker than the intralayer
interaction, which give rise to the two independent plasmon modes of
each layer. Thus we have uncoupled plasmon
modes as
\begin{eqnarray}
\begin{array}{l c c c c c c}
\omega^2_+ = q^2 \alpha_1 (2\pi d^2 P_2(\cos
\theta_0)(\frac{1}{c}-q)+\pi d^2 \sin^2 \theta_0 q )
\\
\\
\omega^2_- = q^2 \alpha_2 (2\pi d^2 P_2(\cos
\theta_0)(\frac{1}{c}-q)+\pi d^2 \sin^2 \theta_0 q ).
\label{Eq:dw}
\end{array}
\end{eqnarray}
The  detailed discussion about the existence of plasmon modes in
this weak coupling limit is the same as that for single layer dipolar
gas. If the two layers have the same density and mass, $\omega^2_+ =
\omega^2_-$, i.e., two plasmon modes have the same dispersion relation.

\begin{figure}
\includegraphics[width=0.99\linewidth]{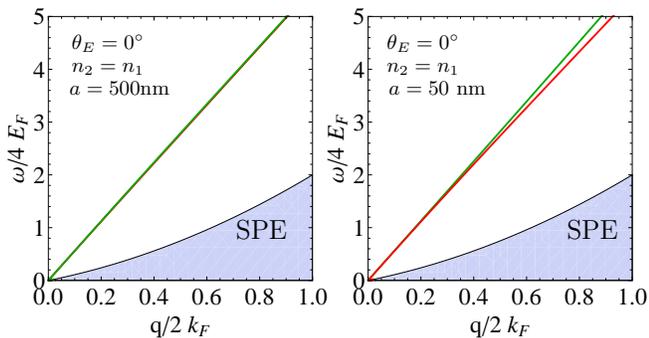}
\caption{(Color online). Calculated plasmon mode dispersions of
  bilayer dipolar system for two different interlayer distances (a)
  $a=0.5 \mu$m and (b) $a=50$ nm at
  $\theta_E = 0^\circ$.
We use $n_1=n_2=10^8$cm$^{-2}$ and $r=4.4$
  and $s = 31$ in this calculation.
The shadowed
  region indicate the single-particle excitation (SPE) region.
}
\label{fig:Diso1}
\end{figure}

\subsection{Numerical results of plasmon modes in bilayer dipolar system}

In this subsection, we show our numerical results of plasmon
modes for the bilayer dipolar system with typical parameters of
$^{40}$K$^{87}$Rb. In Fig.~\ref{fig:Diso1} we show the
calculated plasmon dispersions for an external electric field
perpendicular to $x$-$y$ plane ($\theta_E=0$)
with equal densities of $n_1=n_2 = 10^8 $cm$^{-2}$.
When the separation of two layers are much larger than the inverse of
the Fermi wave vector (i.e. weak coupling limit), the interlayer interaction
is negligible and the bilayer dipolar system
behave like two independent single layers.
Thus, two plasmon modes are almost degenerated and
there is only one plasmon mode showed up in the left panel of
Fig.\ref{fig:Diso1}. When the two
layers are getting closer, the interlayer interaction become
stronger. As a consequence, the density fluctuation in each layer is
coupled and the degenerated plasmon mode is split into two plasmon
modes, especially at large wave vectors, as shown in the right panel of
Fig.\ref{fig:Diso1}.

\begin{figure}
\includegraphics[width=0.99\linewidth]{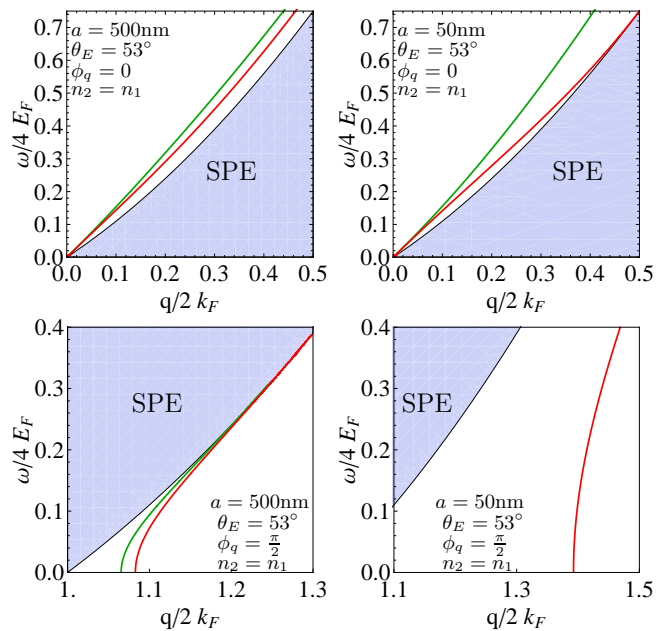}
\caption{(Color online). Calculated plasmon mode dispersions of double
  layer dipolar system for two different interlayer distance and
  $\phi_q$ at $\theta_E = 53^\circ$ with $n_1=n_2=10^8$cm$^{-2}$.  The
  shadowed region indicate the single-particle excitation (SPE)
  region. The solid lines are solutions for $\epsilon(q,\omega)=0$ for
  $r=4.4$ and $s = 31$. }
\label{fig:Diso2}
\end{figure}

In Fig. \ref{fig:Diso2} we show the
calculated plasmon dispersions for $\theta_E=53^\circ$, (i.e.,  the
direction of an external electric field
is close to the critical angle of $\theta_E=55^\circ$).
The plasmon dispersions are calculated with equal densities of
$n_1=n_2 = 10^8$cm$^{-2}$ for
two different interlayer distance ($a=500$nm and 50nm) and two
different directions of the wave vector ($\phi_q=0$ and $\pi/2$)
because of the anisotropic properties of the interlayer
interaction. (Note that for $\theta_E=0$ the plasmon modes are
isotropic and independent of $\phi_q$). Since $\theta_E$ is close to the
critical angle the plasmon modes approach to the upper boundary of
the electron-hole continuum. For $\phi_q=0$ (i.e. the 2D wave vector is along $x$-axis)
both in-phase and out of phase plasmon modes are located above the SPE
region and they are undamped. As long as $\theta_E \ge \bar{\theta}_E$
we find two undamped modes even though the two modes are almost
degenerate for weak coupling limit. When the layer separation is small (i.e., strong coupling limits), the
out-of-phase mode (lower energy mode) gets closer to the SPE
region. For $\theta_E \ge
\bar{\theta}_E$ the out-of-phase mode enters into the SPE and is
overdamped. Thus,  there is only one
undamped mode for $\theta_E \ge \bar{\theta}_E$.
For $\phi_q=\pi/2$ both modes have similar feathers as that for
$\phi_q=0$ but they are
very close to the upper boundary of SPE region. When
the $\theta_E$ is greater than the critical angle we find no plasmon modes
above the upper boundary of SPE region (i.e., $\omega =
\dfrac{q^2}{2 m}+v_F q$). But there are plasmon modes appearing
below the lower boundary of SPE region (i.e., $\omega = {q^2}/{2
  m}-v_F q$) for $\phi_q=\pi/2$ (the lower panel of
Fig.~\ref{fig:Diso2}). As the interlayer coupling gets stronger, the
in-phase mode with higher frequency is pushed
into the SPE region and only the out-of-phase mode with lower frequency survived.

\begin{figure}
\includegraphics[width=0.99\linewidth]{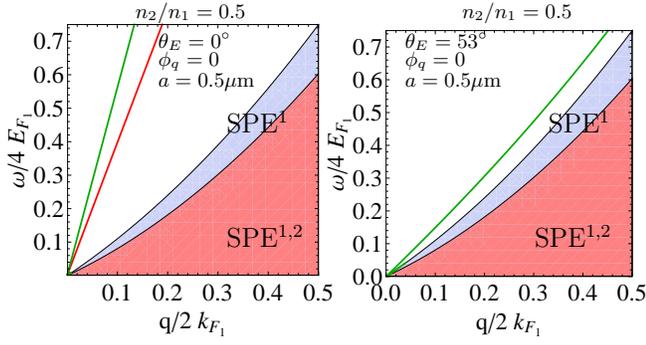}
\caption{(Color online). Calculated plasmon mode dispersions of double
  layer dipolar system for two different $\theta_E$ at $\phi_q =0$
  with $n_1=10^8$cm$^{-2}$ ($k_F = \sqrt{4 \pi n_1}$), $n_2/n_1=0.5$
  and interlayer distance $a=0.5\mu$m.  The shadowed region indicate
  the single-particle excitation (SPE) region. The solid lines are
  solutions for $\epsilon(q,\omega)=0$ for  $r=4.4$ and $s = 31$. The
  left panel is for the electric field perpendicular to the $xy$ plane
  $\theta_E = 0^\circ$, while the right panel is for the electric
  field titled at $\theta_E = 53^\circ$.}
\label{fig:Ddif}
\end{figure}

In Fig. \ref{fig:Ddif}, we calculated plasmon mode dispersions of
bilayer dipolar system with different densities ($n_2/n_1 = 0.5$) for
two different $\theta_E$ at $\phi_q =0$. Fig.~\ref{fig:Ddif}(a) presents the
plasmon modes for the isotropic interaction, i.e.,
the external electric field is perpendicular to the $xy$ plane
($\theta_E = 0^\circ$). For a large interlayer distance ($a =
0.5\mu$m), the coupling between the two layers is very weak and they
behave like two independent layers as shown in Eq.~(\ref{Eq:dw}).  From
Eq.~(\ref{Eq:dw}), we know that, for $\theta_E = 0$ and weak coupling
limit, both plasmon modes have linear dispersion relation and the
slope of the plasmon modes  is proportional to their densities. Thus,
two separated plasmon modes are undamped.
Fig.~\ref{fig:Ddif}(b) shows the
plasmon dispersion of the anisotropic bilayer dipolar system for
$\theta_E = 53^\circ$, $\phi_q=0$ and $a = 0.5\mu$m. There is only one undamped mode $\omega_+$ because
$\omega_-$ lies inside the SPE region.

\begin{figure}
\includegraphics[width=0.98\linewidth]{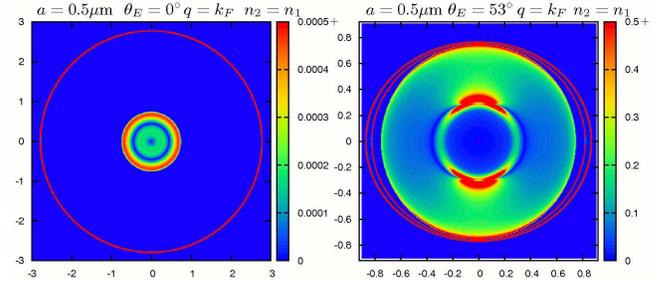}
\caption{(Color online). The density plots of the absolute value of
  the loss function -Im$[\epsilon(k_F,\omega)^{-1}]$ of  bilayer
  dipolar system with same densities in ($\omega, \phi_q$) polar
  coordinate space for fixed density $n_1=n_2=10^8$cm$^{-2}$, $r=4.4$,
  $s = 31$, $a=500$nm and $\theta_E =0^\circ$(left) and $\theta_E =
  53^\circ$(right).}
\label{fig:i}
\end{figure}

\begin{figure}
\includegraphics[width=0.98\linewidth]{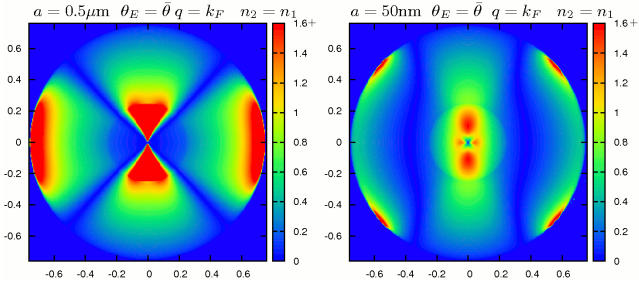}
\caption{(Color online). The density plots of the absolute value of
  the loss function -Im$[\epsilon(k_F,\omega)^{-1}]$ of bilayer
  dipolar system with same densities  in ($\omega, \phi_q$) polar
  coordinate space for fixed density $n_1=n_2=10^8$cm$^{-2}$, $r=4.4$,
  $s = 31$, $\theta_E = \bar{\theta}$ and $h=500$nm(left) and
  $a=50$nm(right).}
\label{fig:ani}
\end{figure}

In Figs.~\ref{fig:i} and \ref{fig:ani} we show the calculated loss
function (i.e., -Im$[\epsilon(k_F,\omega)^{-1}]$) of the bilayer
dipolar system for a fixed wave vector ($q = k_F$).
The loss function is related to the
dynamical structure factor $S(q, \omega)$ by $S(q, \omega)
\propto$-Im$[\epsilon(k_F,\omega)^{-1}]$ and the dynamical structure
factor gives the direct measure of the spectral strength of the
various elementary excitations\cite{Giuliani,Euyheon_PRB09}.
We plot the loss
functions in $(\omega,\phi_q)$ space
to describe the angular dependence of the plasmon
modes. The
loss function is an experimental observable which can be measured with
Raman-scattering spectroscopies. The plasmon modes exist when both
real and imaginary part of the dielectric function equal zero (i.e.,
Re$[\epsilon (q, \omega)]=0$ and Im$[\epsilon (q, \omega)]=0$), which
leads the imaginary part of the inverse dielectric function
-Im$[\epsilon(k_F,\omega)^{-1}]$ to be a $\delta$-function indicating an
undamped plasmon modes. On the other hand, the broadened peak
in the loss function gives the damped plasmon modes. The damping in
the plasmon mode is induced by particle-hole pair creation because we
neglect any disorder effects in this calculation.

In Fig.~\ref{fig:i}(a) we show the result when the applied electric  field
is perpendicular to the plane. In this case
both interlayer and intralayer interaction are isotropic.
The solid circle gives the undamped
plasmon modes, at which the loss function becomes singular. Since
two plasmon modes are degenerate and they are independent of the wave
vector direction $\phi_q$,
the plasmon modes appear as one solid circle in
the ($\omega, \phi_q$) space. For $\theta_E =
53^\circ$, the two plasmon modes can also
clearly be seen in Fig.\ref{fig:i}(b).
For $\theta_E = 53^\circ$, both intralayer and
interlayer interaction are anisotropic, so that the plasmon modes appear as
ellipses in the ($\omega, \phi_q$) space and they
become much closer along the $y$-axis.

In Fig.~\ref{fig:ani}, we present the density plots of
the loss function -Im$[\epsilon(k_F,\omega)^{-1}]$ of bilayer
dipolar system for $\theta_E =  \bar{\theta}$, where the short
distance cut-off disappears from the intralayer interaction and the
bilayer system becomes totally anisotropic. In
Fig.~\ref{fig:ani} the plasmon modes are shown for interlayer distance
(a) $a = 0.5\mu$m (strong coupling limit) and (b) $a =
50$nm (weak coupling limit). There is no $\delta$-function like peak
in the loss function
for $q = k_F$ in both cases, which correspond to damped
plasmon modes.

\section{Plasmon mode in dipolar superlattice}
\label{sec:sup}

In this section, we discuss the plasmon modes of a superlattice system
which consist of the infinite number of parallel and equally separated
two dimensional dipolar system. Due to the long range
behavior of the dipolar interaction, it is necessary to couple all the
layers, which changes the dielectric function in Eq.~(\ref{Eq:sd}) to an
infinite matrix equation\cite{Sarma_PRL09}:
\beq
|\delta_{l,l'}-V_{l,l'}(q)\Pi_{l'} (q,\omega)|=0
\eeq
where $l$ (or $l'$)$= 0, \pm 1, \pm 2$, ..., are the indices of the
layers and each layer is placed parallel to the $xy$ plane with the
position $z = l a$ in the $z$ direction.  $V_{l,l'}(q)$ is the dipolar
interaction between the $l$th and the $l'$th layer, which can be
written as
\begin{equation}
\begin{array}{l l l l l l }
l=l':
\\
V_{l l} = V_0 (q) = \pi d ^2 (3 \cos^2 \theta_E - 1) \dfrac{1}{c}
\\
\ \ \ \ \ \ \ \ \ \ \ \ + 2 d^2 q \pi (\sin ^2 \theta _E\cos ^2 \phi
_q-\cos ^2 \theta _E);
\\
l \neq l':
\\
V_{l l'}(q) =  2 d^2 q \pi  e^{-a q |l-l'|} (\sin ^2 \theta _E\cos ^2
\phi _q-\cos ^2 \theta _E
\\
\ \ \ \ \ \ \ \ \ \ \ \ -i  \text{sgn} (l-l')\sin  2\theta_E\cos  \phi _q ),
\label{Eq:sin}
\end{array}
\end{equation}
where $a$ is the
superlattice period (i.e., the separation between adjacent layers in
the $z$ direction), $\text{sgn} (l-l')$ equals $1 (l>l')$ or $-1
(l<l')$, and $\Pi_{l'} (q,\omega)=\Pi (q,\omega)$ is the irreducible
polarizability of each 2D dipolar system given in Eq.\ref{Eq:p1}.
The difference between intralayer and interlayer interaction in
calculating the dipolar system is that the intralayer interaction in
momentum space has the short distance cut-off $c$, while the
interlayer interaction does not depending on this short distance
cut-off.

The plasmon modes of the infinite periodic system can be calculated
from the self-consistent field method
described in Ref.~[\onlinecite{Sarma_PRB82}]. Since the superlattice
is perfectly periodic in the $z$ direction with periodicity $a$, we
assume the following ansatz $n_l(q,\omega) = n_0 (q, \omega)
e^{i k_z l a}$, where $a$ is the interlayer distance and $l$ is the
layer index. The $k_z$ introduced in the above ansatz labels the
induced density fluctuation in the infinite periodic system, which is
restricted in the first Brillouin zone of the superlattice, i.e., $0
\leq k_z \leq \frac{2\pi}{a}$.
Then the plasmon modes for the dipolar superlattice system are
given by the roots of following equation
\begin{equation}
1- \Pi(q, \omega) V_0(q) - \Pi(q, \omega) \sum _{l'(l\neq l')} V_{l
  l'}(q)e^{-i k_z (l-l')a}=0
\label{Eq:s1}
\end{equation}
where $V_0(q)$ is the intralayer interaction in  momentum space while
$V_{l- l'}(q)$ is the interlayer interaction. With the help of formula
given in Appendix.~\ref{app:superla}, Eq.~(\ref{Eq:s1}) can be explicitly
written as:
\begin{widetext}
\begin{equation}
1- \Pi(q, \omega) V_0(q) - \Pi(q, \omega) 2 d^2 q \pi \big[(\sin ^2
  \theta _E\cos ^2 \phi _q-\cos ^2 \theta _E)S_1 (q, k_z)
-\sin  2\theta_E\cos  \phi _q S_2(q, k_z)\big]=0,
\label{Eq:s3}
\end{equation}
\end{widetext}
where
\beq
S_1 (q,k_z)  =\dfrac{\cos k_z a-\exp{(-a q)}}{\cosh a q -\cos k_z a},
\eeq
and
\beq
S_2 (q,k_z)= \dfrac{\sin k_z a}{\cosh a q -\cos k_z a}.
\eeq

We investigate the long wavelength plasmon modes for the strong
coupling $q a \gg 1$ and the weak coupling
case $q a \ll 1$ analytically in the following two subsections.

\subsection{Strong coupling case $q a \ll 1$}

First, we consider the $k_z \neq 0$ case. With the asymptotic form of
the polarizability $\Pi(q, \omega)\simeq \dfrac{n}{m}
\dfrac{q^2}{\omega^2}$ we can rewrite Eq.~(\ref{Eq:s3}) for $q a \ll 1$
as
\begin{widetext}
\begin{equation}
 1- 2\pi d ^2 \dfrac{n q^2}{m \omega^2} \Big[\dfrac{P_2(\cos
    \theta_E) }{c}+ (\sin ^2 \theta _E\cos ^2 \phi _q-\cos ^2 \theta
  _E)\dfrac{a q^2}{1 -\cos k_z a}
-q \sin  2\theta_E\cos  \phi _q \dfrac{\sin k_z a}{1 -\cos k_z a}\Big] = 0.
\label{Eq:sstron}
\end{equation}
Then we have the plasmon modes for the
superlattice dipolar system in the long length limit
\begin{equation}
\omega^2 \simeq q^2  \dfrac{2 n \pi d ^2}{m }\Big[\dfrac{P_2(\cos
    \theta_E) }{c}+ (\sin ^2 \theta _E\cos ^2 \phi _q-\cos ^2 \theta
  _E)\dfrac{a q^2}{1 -\cos k_z a}\Big].
\label{Eq:ss2}
\end{equation}
\end{widetext}
The plasmon mode in superlattice system is strongly dependent on the
direction of the external
electric field $\theta_E$ and the wave vector direction $\phi_q$ as
well as the wave vector $k_z$ that labels the density fluctuation. When the
electric field is perpendicular to the plane,
we find that the plasmon mode has
linear dispersion relation $\omega \propto q$.
For $\theta_E = \bar{\theta}$, in which the interactions are anisotropic,
the plasmon mode dispersion becomes
\begin{equation}
\omega \simeq q ^{\frac{3}{2}} \sqrt{\dfrac{2 n \pi d ^2}{3 m
  }\big(\dfrac{a q \cos 2\phi_q}{1-\cos k_z a}-2 \sqrt{2} \frac{\cos
    \phi_q \sin k_z a}{1- \cos k_z a} \big)} .
\end{equation}
In order for this mode to lie above the SPE region, i.e. $\omega > q
v_F$,  it is required to be $r \gg 1$.

For $k_z = 0$, the long wavelength plasmon mode can be calculated
from  Eq.~(\ref{Eq:s3}) to be
\begin{equation}
\omega^2 \simeq   \dfrac{q^2 2 n \pi d ^2}{m}\Big[\dfrac{P_2(\cos
    \theta_E) }{c}+ \dfrac{2(\sin ^2 \theta _E\cos ^2 \phi _q-\cos ^2
    \theta _E)}{a}\Big].
\label{Eq:ssl}
\end{equation}
In particular, the plasmon dispersion for $\theta_E =0$ becomes
\begin{equation}
\omega \simeq q \sqrt{\dfrac{2 n \pi d ^2 (a-2c)}{m a c} }.
\label{Eq:ssi}
\end{equation}
From Eq.~(\ref{Eq:ssi}) it is clear that only for $a>2c$ there is
well defined plasmon mode with
linear dispersion relation. For the total anisotropic case
($\theta_E=\bar{\theta}_E$),
the plasmon mode also has linear dispersion relation which can written as:
\begin{equation}
\omega \simeq q \sqrt{\dfrac{4 n \pi d ^2 \cos2\phi_q}{3m a} }.
\label{Eq:ssani}
\end{equation}
Thus the plasmon mode only exists for certain angles such that
$\cos2\phi_q>0$.

Compared to the dipolar superlattice system, the plasma frequency for the superlattice system with Coulomb interaction can be written as\cite{Sarma_PRL09}:
\beq
\omega (q \rightarrow 0; k_z =0)=\Big(\dfrac{4\pi e^2 n}{a \kappa m}\Big)^{1/2}
\eeq
which has the precise character of the corresponding three dimensional plasmon (with a finite gap) in the long wavelength limit.

\subsection{Weak Coupling case $q a \gg 1$}

For the weak coupling ($q a \gg 1$)
Eq.~(\ref{Eq:s3}) becomes
\beq
1- \Pi(q, \omega) V_0(q) =0
\eeq
for all values of $k_z$. Thus
the long wavelength plasmon mode of superlattice system becomes the mode
of the single layer dipolar system as
discussed in Sec.\ref{sec:single}. Each layer has its
own plasmon mode in the weak coupling limits as expected.

\section{Summary and Conclusions}

In summary, we have derived the collective plasmon modes in 2D dipolar Fermi liquids,
and also considered spatially separated bilayer and superlattice
dipolar systems  within the random phase approximation, which is valid
for weakly interacting system. We  have also
calculated the loss function for bilayer fermionic dipolar gas, which
could be measured in experiments. The 2D dipolar interaction with both
s and d-wave components in the wave vector space leads to several
unexpected features in the plasmon modes such as undamped mode if
dipole along the $z$-axis and the anisotropic plasmon dispersion
relation if the dipole along other direction than the $z$-axis. Our
predicted plasmon modes is clearly distinguished from the extensively
studied two dimensional electron system.

Future work ought to include higher order corrections to the
polarizability and the finite temperature corrections, yet this must
await the successful fabrication of quasi two dimensional dilute
fermionic dipolar gases. We note that our theory for the collective mode spectra of 2D dipolar Fermi
liquids should have considerable relevance for the recently made polar molecular gases Ref.~[\onlinecite{Ospelkaus_NP08,Ni_Sci08,Ni_PCCP09,Ospelkaus_PRL10,Griesmaier_PRL77,Ospelkaus_FD09}]. At low enough
temperatures, $T \ll T_F$ where $T_F$ is the Fermi temperature of the dipolar system, our theory should in principle describe the
laboratory polar molecular fermionic systems, and the excitation spectra of the dipolar system should have clear
signatures of the collective mode spectra presented in our work.

\begin{acknowledgments}
QL acknowledges helpful discussions with Kai Sun. The work is
supported by AFOSR-MURI and NSF-JQI-PFC.

\end{acknowledgments}

\appendix
\section{}
\label{app:ftinterlayer}
Below we provide the detailed calculation of the interlayer dipolar
interaction in the wave vector space.
In the real space the interlayer dipolar interaction becomes
\begin{equation}
V_{12}(\vec{r})=\frac{d^2}{(r^2+a^2)^{\frac{3}{2}}}\left[1-\frac{3(r
    \cos\phi \sin\theta_E+h \cos\theta_E)^2}{r^2+h^2}\right],
\end{equation}
where $\phi_q$ is the angle between momentum $q$ and the $x$-axis.
We can devide above equation into three different
partss depending on the symmetry
\begin{equation}
V^{(a)}_{12}(r)=\frac{d^2}{(r^2+a^2)^{\frac{3}{2}}}(1- \frac{3a^2 \cos^2
  \theta_0}{r^2+h^2}),
\end{equation}
\begin{equation}
V^{(b)}_{12}(r)=\frac{d^2}{(r^2+a^2)^{\frac{5}{2}}}(3 r^2 \cos^2\phi
\sin^2 \theta_0),
\end{equation}
and
\begin{equation}
V^{(c)}_{12}(r)=\frac{d^2}{(r^2+a^2)^{\frac{5}{2}}}(6 r \cos\phi
\sin\theta_0 a \cos \theta_0),
\end{equation}
where $V_{12}(q)=V^{(a)}_{12}(q)-V^{(b)}_{12}(q)-V^{(c)}_{12}(q)$.
Then we have
\begin{eqnarray}
\begin{array}{l c c c c c c}
V^{(a)}_{12}(q) = d^2 \int _0^{\infty } \int _0^{2 \pi }
\dfrac{1-3\frac{a^2
    \cos^2\theta_0}{a^2+r^2}}{(a^2+r^2)^{\frac{3}{2}}}\exp(i
\vec{q}\cdot\vec{r})r dr d\phi
\\
\\
\ \ \ \ \ \ = 2 q \pi d^2 e^{-a q}(\dfrac{\sin^2\theta_0}{a q}-\cos^2\theta_0),
\end{array}
\end{eqnarray}
where $J_0(x)$ is the Bessel function of the first kind and
$\vec{q}\cdot\vec{r}= q r \cos(\phi-\phi_q)$.
\begin{eqnarray}
\begin{array}{l c c c c c c c }
V^{(b)}_{12}(q) = d^2 \int _0^{\infty } \int _0^{2 \pi } \dfrac{3 r^2
  \cos^2 \phi \sin^2 \theta_0}{(a^2+r^2)^{\frac{5}{2}}}\exp(i
\vec{q}\cdot\vec{r})r dr d\phi
\\
\\
= 2 d^2 q \pi  e^{-a q}\sin ^2\theta _0(\dfrac{1}{a q}-\cos^2\phi_q ),
\end{array}
\end{eqnarray}
and
\begin{eqnarray}
\begin{array}{l c c c c c c}
V^{(c)}_{12}(q) = d^2 \int _0^{\infty } \int _0^{2 \pi } \dfrac{3
  r^2 a \cos \phi \sin 2\theta_0}{(a^2+r^2)^{\frac{5}{2}}}\exp(i
\vec{q}\cdot\vec{r})dr d\phi
\\
\\
= 2 d^2 q \pi  e^{-a q}\sin  2\theta _0\cos\phi_q \ i
\end{array}
\end{eqnarray}
which is pure imaginary number.
Finally we have the interlayer interaction in momentum space
\beq
V_{12}(q) =  2 d^2 q \pi  e^{-a q} (\sin ^2 \theta _E\cos ^2 \phi
_q-\cos ^2 \theta _E-\sin  2\theta_E\cos  \phi _q \ i).
\eeq

We also find $V_{21}(q)$ from $V_{21}(\vec{r})$
\begin{equation}
V_{21}(\vec{r})=\frac{d^2}{(r^2+a^2)^{\frac{3}{2}}}\left[1-\frac{3(r
    \cos\phi \sin\theta_E-h \cos\theta_E)^2}{r^2+h^2}\right]
\end{equation}
the interlayer dipolar interaction $V_{21} (q)$ in the wave vector
space is given the complex conjugate of $V_{12} (q)$, which can written as:
\beq
V_{21}(q) =  2 d^2 q \pi  e^{-a q} (\sin ^2 \theta _E\cos ^2 \phi
_q-\cos ^2 \theta _E + \sin  2\theta_E\cos  \phi _q \ i)
\eeq

\section{}
\label{app:superla}
Below we provide detailed calculation of two infinite sums in Sec. \ref{sec:sup}.
\begin{eqnarray}
S_1 (q, k_z) & = &\sum_{l\neq 0} \text{exp} [-|l| a q -i k_z l a]
\nonumber \\
& = & 2 \text{Re} [\dfrac{\text{exp}(-l a q-i k_z l a)}{1-\text{exp}(-l a
    q-i k_z l a)}] \nonumber \\
& = & \dfrac{\cos k_z a-\exp{(-a q)}}{\cosh a q -\cos k_z a},
\end{eqnarray}
and
\begin{eqnarray}
S_2 (q, k_z) & = & -i \sum_{l\neq 0} \text{sgn}(l)\text{exp} [-|l| a q
  +i k_z l a] \nonumber \\
& = & 2  \text{Im} [\dfrac{\text{exp}(-l a q+i k_z l a)}{1-\text{exp}(-l a
    q+i k_z l a)}] \nonumber \\
& = & \dfrac{\sin k_z a}{\cosh a q -\cos k_z a}.
\end{eqnarray}

\end{document}